\documentclass{mn2e}
\usepackage{epsf,times}
\usepackage{amsmath}
\usepackage{graphicx}
\usepackage{color}
\newcommand{\etal}{{et al}\/.}
\begin{document}

\title[X-ray jets in FR-I radio galaxies]{What determines the properties of the X-ray jets in FR-I radio galaxies?}

\author[J.J.~Harwood \& M.J.~Hardcastle]{Jeremy J.\ Harwood\thanks{E-mail: jeremy.harwood@physics.org} \& Martin J.\ Hardcastle\thanks{E-mail:m.j.hardcastle@herts.ac.uk}\\School of Physics, Astronomy and Mathematics, University of Hertfordshire, College Lane, Hatfield, Hertfordshire AL10 9AB}

\maketitle

\begin{abstract}

We present the first large sample investigation of the properties of jets in Fanaroff and Riley type I radio galaxies (FR-I) based on data from the \emph{Chandra} archive. We explore relations between the properties of the jets and the properties of host galaxies in which they reside. We find previously unknown correlations to exist, relating photon index, volume emissivity, jet volume and luminosity, and find that the previously long held assumption of a relationship between luminosities at radio and X-ray wavelengths is linear in nature when \emph{bona fide} FR-I radio galaxies are considered. In addition, we attempt to constrain properties which may play a key role in determination of the diffuse emission process. We test a simple model in which large-scale magnetic field variations are primarily responsible for determining jet properties; however, we find that this model is inconsistent with our best estimates of the relative magnetic field strength in our sample. Models of particle acceleration should attempt to account for our results if they are to describe FR-I jets accurately.

\end{abstract}


\section{Introduction}
\label{intro}
\subsection{X-ray jets in radio galaxies}

With the launch of the \emph{Chandra} X-Ray observatory in 1999, it quickly became apparent (e.g. Worrall \etal\ 2001a; Sambruna \etal\ 2004; Marshall \etal\ 2005) that emission at X-ray wavelengths was a common, if not a universal (Hardcastle \etal\ 2002) feature of the jets in Fanaroff \& Riley (1974) class I radio galaxies (FR-I). It is generally argued that the emission of X-rays in most FR-Is, where the radio emission is brightest at the central and inner regions, can be well modelled as synchrotron emission (e.g. Hardcastle, Birkinshaw \& Worrall 2001; Wilson \& Yang 2002; Marshall \etal\ 2002; Brunetti \etal\ 2003 ; Perlman \& Wilson 2005; Harris \& Krawczynski 2006; Massaro, Harris \& Cheung 2011). In many cases, a `one-zone' model in which emission originates from a single electron population is used (e.g. Laing \etal\ 2008) and this is supported by observations of smooth spectral transitions between the radio and optical components of jets to those observed at X-ray wavelengths (e.g. Hardcastle \etal\ 2001). In reality, although a one-zone model provides good agreement with many of the observed jet properties, it is unlikely that it provides a full description of the mechanism by which these jets radiate. For example, both Perlman \& Wilson (2005) for M87 and Hardcastle \etal\ (2003) for Centaurus A require variations in the electron energy spectrum as a function of position. The short emitting lifetime of particles radiating at X-ray frequencies (see below) relative to the size of jets observed in FR-I galaxies also means that it is improbable that the electron energy spectrum will be the same at all points along the jet. There is therefore no \emph{a priori} reason to assume a one-zone model. That said, this simplified model does seem to provide a good description of many observed jet properties and so is commonly used. However, while a synchrotron model appears to describe the jets well, it leaves open the very important question of how the emitting electrons are accelerated to the high energies required to produce the X-rays.

In the synchrotron process, the lifetime over which a particle will radiate depends on the Lorentz factor of the emitting particle and the magnetic field in which it resides (e.g. Longair 1994). For emission in an FR-I jet at 1 keV in a 10 nT magnetic field, and assuming a magnetic field that is approximately in equipartition with the radio emissions (e.g. Leahy 1991), the expected radiative lifetime of an electron is expected to be only of the order of hundreds of years. If we compare this to the approximate size of the X-ray jet of the FR-I radio galaxy Centaurus A, which extends to a distance of approximately 4.5 kiloparsecs from the active galactic nucleus (AGN), it would take a particle travelling at the speed of light around 15000 years to travel from the central region to its furthest extent (see Burns, Feigelson \& Schreier 1983; Harris \& Krawczynski 2006; Hardcastle \etal\ 2007 and references therein for similar arguments). The X-ray emission cannot therefore be due to particles accelerated near the AGN; instead the emitting particles must be accelerated \emph{in situ} in the jet (e.g. Biretta \& Meisenheimer 1993; Brunetti \etal\ 2003; Hardcastle \etal\ 2007; Goodger \etal\ 2009). Various theoretical models exist for the cause of this diffuse \emph{in situ} acceleration, such as second order Fermi acceleration (Stawarz \& Ostrowski 2002; Rieger \etal\ 2007), magnetic reconnection (Birk \& Lesch 2000) and decaying beams of ultra high-energy neutrons (Atoyan \& Dermer 2004); however, determining which, if any, of these models if correct is fraught with difficulties.

Investigations into the mechanisms involved in the acceleration of jet particles to high energies have so far mainly focused around the details of individual galaxies  (e.g. Kataoka \etal\ 2006 and Goodger \etal\ 2009 for Centaurus A; Birk \& Lesch 2000, Perlman \& Wilson 2005 and Sahayanathan 2008 for M87; see references in Table 1 for an extensive list of individual studies). However, with no large scale sample available for comparison, it has been hard to determine whether a feature of a particular galaxy is common across all FR-I galaxies, or whether it is unique to that target. This may, in part at least, be due to the limited number of objects which have been available for analysis; however, with a growing number of galaxies in which X-ray jets are known to be present, enough data now exist for investigations to be made into models requiring empirical data across a wider range of targets. When dealing with these bulk relations and treating the jets as a whole in this manner, we must be careful in our consideration of the underlying physics. We know for example that these jets must have at least two emission mechanisms; one for the small scale structure of features such as knots, and another for the large scale diffuse emission. All of these small scale and localised factors will introduce scatter in to any relations found. However, it is important to understand the large scale relations if we are to understand the underlying physics of these jets as a whole. Any general relations predicting the jets' photon indices \footnote{We define the photon index to be $\Gamma \equiv \alpha +1$  where the spectral index $\alpha$ relates the flux, $S$, and frequency, $\nu$, by $S \propto \nu^{-\alpha}$}, or relating jet properties to the intrinsic properties of the host galaxy in which the jets reside, have yet to be explored. It is critical that any acceleration model put forward should account for relations found in these populations. In this paper we therefore aim to explore whether any relations exist, on large scales, between a range of jet  properties and the properties of the host galaxy in which they reside, in an attempt to locate the key variables upon which particle acceleration is dependent. In doing so we hope to give further insight into the acceleration process, and to provide a benchmark against which comparison of investigations of individual galaxies can be made.

The data we collect also allows us to test those relations which are currently assumed to exist. For example, as it is generally accepted that the emission mechanism of FR-I jets is synchrotron dominated at all wavelengths (see above), we would expect to see a smooth transition between spectra at varying wavelengths. There is therefore a general expectation that the luminosity of a galaxy's radio jets will be related to that in the X-ray; however, there is currently little empirical support at large scales for this prediction. As it is almost a requirement for this condition to be true to model X-ray jets in FR-I galaxies in a one-zone synchrotron framework, it is important to test its validity.

We make use of data collected from the \emph{Chandra} archive to determine the luminosities and photon indices for a large sample of known X-ray jets in FR-I galaxies. These properties, along with the derived ratio of X-ray to 8.46 GHz radio flux and jet emissivity, are then compared to the luminosity of the host galaxy across the ultraviolet, visual and infrared pass bands as well as the high frequency radio luminosity in the region containing the X-ray component, the low frequency luminosity of the radio galaxy as a whole and the luminosity of the AGN core. Throughout this paper, a concordance model in which $H_0=71$ km s$^{-1}$ Mpc$^{-1}$, $\Omega _m =0.27$ and $\Omega _\Lambda =0.73$ is used. In Section 2, details of our target selection and data reduction methods are discussed. In Section 3 we present our statistical analysis of the results, highlighting the correlations which exist between jet properties. Finally, in Section 4 we discuss the implications of these results for current theory and test a simple acceleration model.

\section{Target Galaxies and Analysis}
\label{analysis}

We made use of the XJET website\footnote{http://hea-www.harvard.edu/XJET} (X-ray emission from extragalaxtic radio jets; compiled by Massaro, Harris \& Cheung), which acts as a clearing house for radio galaxies and quasars in which X-ray emission has been detected and which are associated with the radio jet component, to compile a list of possible targets. As discussed in Section 1, the mechanism by which FR-I and intermediate FR-I/IIs emit is generally agreed to be synchrotron. This is in contrast to the on-going discussions as to the emission mechanism of FR-II galaxies in which no consensus has yet been reached. To maximize the number of data points available, all FR-I, intermediate FR-I/FR-II and BL Lac type galaxies were selected from the XJET list as of August 2010 for further investigation. The well known source 3C 78 was also included as a known good example of its class (e.g. Trussoni \etal\ 1999). For a galaxy within this list to be considered a suitable target, it must also have satisfied the following criteria;

\begin{enumerate}

\item It should be a radio galaxy of type FR-I or be classed as intermediate between FR-I and FR-II.
\item Data should be available in the \emph{Chandra} archive, and the count rate in the jet should have a fractional error $\le$ 0.3 (equivalent to $>$ 11 counts within the jet region).
\item Radio images of the target comparable in resolution to the \emph{Chandra} images should be available.

\end{enumerate}

Those galaxies which fulfilled these criteria were then divided into two subsamples; one complete sample including all BL Lacs and intermediate FR-I / FR-II cases and one containing only galaxies which can be considered \emph{bona fide} FR-Is. A detailed listing of which galaxies are included is found in Table 1.

 The \emph{Chandra} search and retrieval tool\footnote{http://cxc.harvard.edu/cda/chaser.html} (ChaSeR) was then used to resolve the target coordinates via the Simbad astronomical database\footnote{http://simbad.u-strasbg.fr/simbad/} and retrieve a list of available data from the \emph{Chandra} archive. In most cases only a limited number of observations were available, but for the best studied objects exposures on the AXAF (Advanced X-ray Astronomy Facility) CCD Imaging Spectrometer (ACIS) with no gratings were used for consistency with the most common setup for the fainter, less studied objects.

We reprocessed all the data using CIAO version 4.2 and the resultant calibrated event files were then used for all subsequent analysis. A summary of all potential target galaxies is displayed in Table 1, along with features noted on the XJET website and any reason for which a galaxy was excluded from further consideration in this paper.

\begin{table*}
\caption{List of target galaxies, observation IDs and noted features}
\label{observations}
\begin{tabular}{lccccccl}
\hline
\hline

Name&Class&OBSID&Exposure (ks)&X-Ray Features&$z$&References&Sample\\
\hline
0313-192&FRI RG&4874&19.17&Inner jet&0.0670&1&Both\\
3C 129&FRI RG&2218 - 2219&41.61&Two inner knots&0.0208&2&Both\\
3C 15&FRI/II RG&2178&28.18&Knot and lobes&0.0730&3&Full only\\
3C 264&FRI&4916&38.33&Jet&0.0217&4&Both\\
3C 296&FRI RG&3968&50.08&Inner jet&0.0237&5&Both\\
3C 31&FRI RG&2147&44.98&Inner 8'' jet&0.0167&6&Both\\
3C 346&FRI/II RG&3129&46.69&Knot&0.1610&7&Full only\\
3C 371&BL Lac (FRI/II)&841&10.25&Inner and outer knots&0.0510&8, 33&Full only\\
&&2959&40.86&\\
3C 465&FRI RG&4816&50.16&Knots&0.0293&9&Both\\
3C 66B&FRI RG&828&45.17&Inner 8'' jet&0.0215&10&Both\\
3C 78&FRI&3128&5.23&&0.0287&11&Both\\
&&4157&53.47&\\
3C 83.1&FRI RG&3237&95.14&E and W knots&0.0251&12&Both\\
4C 29.30&FRI/II RG&2135&8.48&Hotspots&0.0640&13&Full only\\
&&11688 - 11689&201.64&\\
&&12106&30.45&\\
&&12119&56.18&\\
B2 0206+35&FRI RG&856&8.57&Inner jet&0.0369&14&Both\\
B2 0755+37&FRI RG&858&8.26&4'' long jet&0.0428&14&Both\\
Centaurus A&FRI RG&962&36.97&Knotty jet&0.0018&15 - 19&Both\\
&&1600 - 1601&99.61&\\
&&2978&45.18&\\
&&3965&50.17&\\
&&7797 - 7800&286.25&\\
&&8489 - 8490&190.86&\\
&&10722 - 10726&70.55&\\
&&11846 - 11847&9.8&\\
Centaurus B&FRI RG&3120&5.16&Inner jet&0.0130&20&None\\
M84&FRI RG&803&28.85&Two inner knots&0.0035&21&Both\\
&&5908&46.68&\\
&&6131&41.45&\\
M87&FRI RG&241&38.53&Knotty jet&0.0043&22 - 28&Both\\
&&352&38.16&\\
&&1808&14.17&\\
&&2707&99.93&\\
&&3084 - 3088&26.28&\\
&&3717&20.83&\\
&&3975 - 3982&43.01&\\
&&4917 - 4923&37.03&\\
&&5737 - 5748&62.57&\\
 &&6299 - 6305&35.7&& &\\
&&7348 - 7351&20.48&\\
NGC315&FRI RG&4156&56.17&Inner jet knots&0.0165&29&Both\\
&FRI RG&855&5.15&\\
NGC4261&FRI RG&9569&102.24&Inner jet&0.0074&30&Both\\
&FRI RG&834&35.18&\\
NGC6251&FRI RG&847&37.44&Knots&0.0249&31&Both\\
&&4130&49.17&\\
PKS0851+202&BL Lac&9182&49.97&Knots&0.306&32&Full only\\
PKS2201+044&BL Lac&2960&40.4&Inner and outer jet&0.027&33&Full only\\
S5 2007+777&BL Lac (FRI/II)&5709&36.87&Jet&0.342&34&Full only\\
\hline

\end{tabular}

\vskip 5pt
\begin{minipage}{17.5cm}
`Name' lists the name of the galaxy as used within this paper. Centaurus A and B refer to the common names of NGC 5128 and PKS 1343-60 respectively and 0313-192 to OL97 0313-192. `Class' lists the classification of the galaxy, `OBSID' the \emph{Chandra} observation ID. Exposure denotes the total observation time in kiloseconds and `$z$' lists the redshift as stated on the XJET website. `Sample' list which of the samples the galaxies are included in. `Full only' refers to sample containing all FR-Is, intermediate FR-I/IIs and BL Lacs, those stated as `Both' are included in this sample and the sample containing only \emph{bona fide} FR-Is. Note that Centaurus B is excluded from either sample on the basis of it's low Chandra X-ray count and no high resolution VLA maps being available due to its location. Values taken from the XJET website herein are adopted from those declared in previous papers as listed in the `References' column. The references are as follows: 1, Keel \etal\ (2006); 2, Harris \etal\ (2002); 3, Kataoka \etal\ (2003); 4, Perlman \etal\ (2010); 5, Hardcastle \etal\ (2005); 6, Hardcastle \etal\ (2002); 7, Worrall \& Birkinshaw (2005); 8, Pesce \etal\ (2001); 9, Hardcastle \etal\ (2005b); 10, Hardcastle \etal\ (2001); 11, Trussoni \etal\ (1999); 12, Sun \etal\ (2005); 13, Sambruna \etal\ (2004); 14, Worrall \etal\ (2001b); 15, Feigelson \etal\ (1981); 16, Kraft \etal\ (2000);17, Kraft \etal\ (2002); 18, Kraft \etal\ (2003); 19, Hardcastle \etal\ (2003); 20, Marshall \etal\ (2005); 21, Harris \etal\ (2002a); 22, Biretta \etal\ (1991); 23, Marshall \etal\ (2002); 24, Wilson \& Yang (2002); 25, Harris \etal\ (2003); 26, Perlman \etal\ (2003); 27 Perlman \& Wilson (2005); 28, Harris \etal\ (2006a); 29, Worrall \etal\ (2003); 30, Chiaberge \etal\ (2003); 31, Mack \etal\ (1997); 32, Marscher \& Jorstad (2011); 33, Sambruna \etal\ (2007); 34, Sambruna \etal\ (2008).

\end{minipage}

\end{table*}

\begin{figure*}

\centering
\includegraphics[width=8cm]{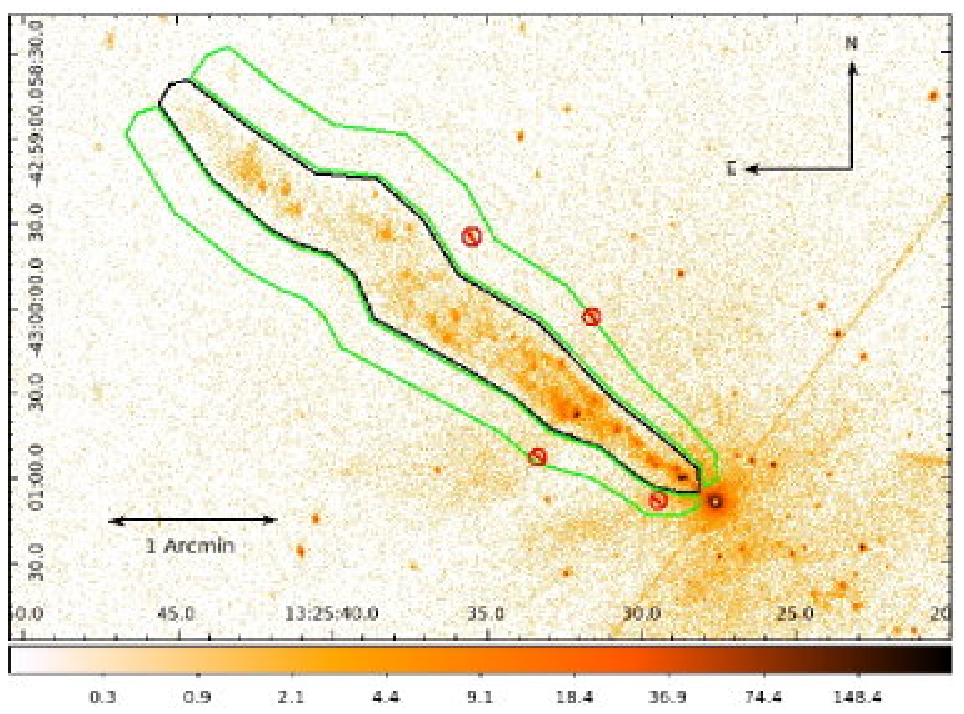}  
\includegraphics[width=8cm]{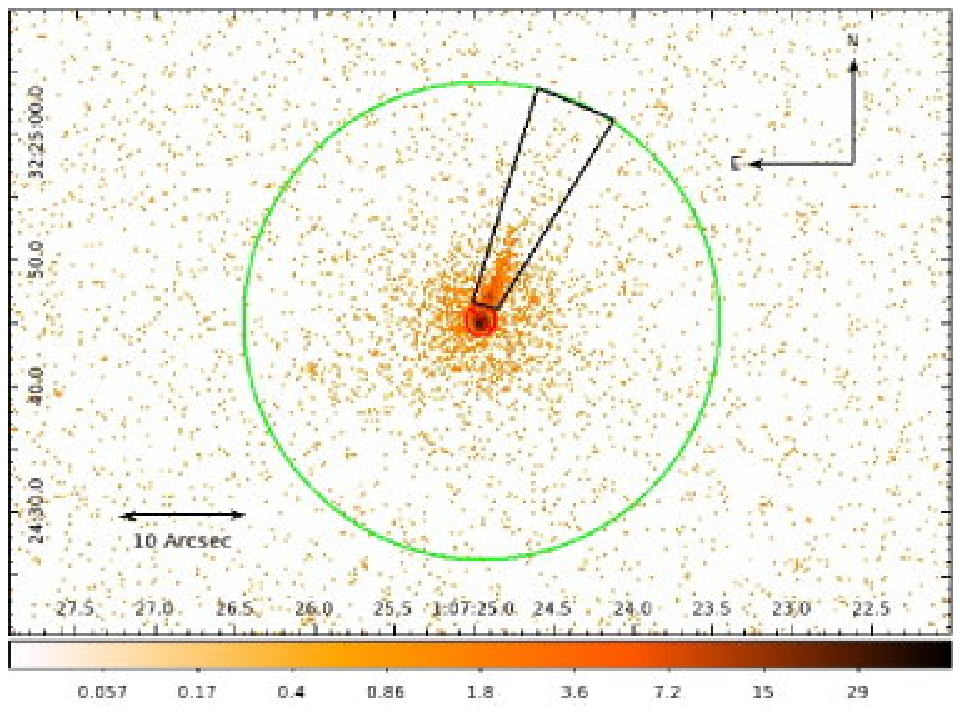}\\
\caption{Examples of background region selection}
\label{bgexamples}

\begin{minipage}{17.5cm}
Solid black regions denote the selected jet region; Solid green denotes are the boundary for inclusion in the background region; Red crossed areas denote regions excluded from consideration. Note: In the case of method 2 (right) black regions are included for the jet, but excluded for background measurements.
The left image displays Cen A as an example of the method used for cases in which background emission varies irregularly as a function of position along the jet. The regions run alongside the jet to its furthest extent. Note the dust lane (see Section 2.3) which runs across the jet approximately $1/3$ of the way from the galactic centre and the exclusion of emission due to point sources which fall in the line of sight.
The right image shows 3C 31 as a case in which background emission is symmetrical about the galactic centre. A large circular region encompasses the largest possible area for an improved uncertainty measurement. Note the exclusion of the jet and AGN regions.
\end{minipage}

\end{figure*}

\subsection{Jet and background region selection}

We used the calibrated images to determine the regions in which the X-ray jets reside and to choose suitable regions for background emission in each data set. For this purpose SAOImage DS9\footnote{http://hea-www.harvard.edu/RD/ds9/} (DS9) was used.

Definition of a jet region for this type of sample can be problematic. Methods such as signal to noise ratio or the use of algorithms to locate the point at which a jet terminates are not suitable due to most jets having a small angular size and low count rate, especially at the jet's furthest extent. Fixed regions both in terms of angular and physical sizes are also not suitable due to the wide variation in both redshift and morphology of the jets, so that there is no `one size fits all' model which can be used for the sample. We therefore followed common practice in that the limit of the jet was defined to be the point at which the emission was no longer discernible by eye from the background emission. This leads to some additional uncertainty and makes the process subjective in nature, but it allows us to retain a sample that is not restricted to one specific size or morphology. To help in reducing the subjective effects of this selection technique the jets were defined without reference to other work and before any measurements had been taken, so as to keep our definition independent of external influences. Once these regions had been defined they were fixed for the duration of the investigation.

Once the jet regions were determined, it was necessary to select an area for which a representative average background count could be taken to allow for any emission from the galaxy, such as thermal emission, that was not representative of the X-ray jet. These were defined using two methods, dependent on the nature of the region in question. For targets in which the background was not uniformly distributed, the region was selected to be approximately parallel to, and comparable in length and shape with, the X-ray jet region so as to account for any changes in background properties along the length of the region (Figure 1, left panel). This was our standard method for targets in which background emission displays significant irregular variation as a function of position along the jet. A good example of this is Centaurus A, in which the obscuring dust lane causes the background emission to vary along the length of the jet.

The second method was used for galaxies in which background emission was on average evenly distributed at any angle from the galactic centre. In these cases, a circular region was taken around the central point to the furthest extent of the jet (Figure 1, right panel). The central AGN, X-ray jet region and any point sources were then excluded from the selection. Although this is a more complex process than the first method, by encompassing a large area in this manner a more accurate average value for the background level can be obtained.

\begin{table*}
\caption{Tested host galaxy properties and their source.}
\label{tested}
\begin{tabular}{ccccc}
\hline
\hline
Tested Property&Region&Survey Name&References&Abbreviation / Comments \\
\hline
B band magnitude&Whole source&Updated Zwicky catalogue&1&UZC\\
H band magnitude&Whole source&The Two Micron All Sky Survey&2&2MASS\\
J band magnitude&Whole source&The Two Micron All Sky Survey&2&2MASS\\
K band magnitude&Whole source&The Two Micron All Sky Survey&2&2MASS\\
4.65 GHz emission&Whole source&New catalogue of 4.85 GHz sources&3&New cat of 4.85 GHz sources\\
8.46 GHz emission&X-Ray jet region&-&-&Measured from radio maps\\
Jet volume&X-Ray Jet Region&-&-&Measured from radio maps\\
Ratio of 4.65 GHz emission&X-ray flux / 4.65 GHz flux&-&-&Derived\\
Acceleration efficiency&X-ray flux / 8.46 GHz flux&-&-&Derived\\
Emissivity&Luminosity / Volume&-&-&Derived\\
Core luminosity&AGN core&-&4 - 5&5 GHz AGN core\\
Core prominence&Core flux / 4.65 GHz flux&-&-&Measure of beaming\\

\hline

\end{tabular}

\vskip 5pt
\begin{minipage}{17.5cm}
`Tested Property' lists the host galaxy property against which the X-ray luminosity and photon index were compared. `Region' lists the area over which this value was measured where `Whole Source' refers to the integrated flux of the host galaxy taken directly from the survey referenced. `Survey Name' lists the source from which the data was taken . The `References' column refers to: 1, Emilio \etal\ (1999); 2, Skrutskie \etal\ (2006); 3, Becker \etal\ (1991); 4, Hardcastle \etal\ (2006); 5, Hardcastle \etal\ (2009). `Abbreviation / Comments' lists the short name used for the survey within this paper where applicable, and any comments on the nature of the property. Derived refers to values which have been calculated from previously stated measurements.
\end{minipage}

\end{table*}

\subsection{Collection of spectral and comparative data}

The spectrum of the jet for each retrieved observation was extracted by use of a Perl script, which again calls upon the CIAO data analysis system by use of the \emph{specextract} command. This script creates a variety of additional files containing a PHA (pulse height amplitude) formatted header compatible with the spectral analysis software XSPEC\footnote{http://heasarc.gsfc.nasa.gov/docs/xanadu/xspec/} by referencing the jet and background region files previously created. The extracted spectra files were loaded into XSPEC and any channels marked as bad due to the binning process were set to be ignored. The column density for each target was obtained using the Colden neutral hydrogen density calculator\footnote{http://cxc.harvard.edu/toolkit/colden.jsp}. The spectral energy distribution (SED) was then fitted in the 0.4 - 7.0 keV range with an absorbed power law model to determine the normalization, photon index and their associated errors. In cases where the target galaxy had multiple observations, spectra were extracted for each and the model was fitted to all the spectra simultaneously. A flux for the jet was then derived from the normalization values which was converted to a luminosity in the standard manner.

Due to the faint nature of the jets and the scarcity of target galaxies, every effort was made to include as much data as possible in the sample. In some cases, the number of counts within the jet region was too low for any usable data to be obtained. However, in some intermediate cases where the counts were too low to extract a reliable SED automatically via XSPEC, so that a photon index could not be measured, it was still possible to obtain a value for the flux by extrapolation of the available data points. In these cases, we used non-grouped versions of the spectra. These files were loaded into XSPEC and energies between 0.0 and 0.4 keV and above 7.0 keV were ignored. An average photon index of 2.1 was taken from the jets in which fits were possible (Table 3) and, along with the column density, entered into the model fixing these values to leave the normalization as the only variable. Using the model in this way, it was possible to obtain the normalization of a power-law model that would produce the observed net count rate from which a flux can be derived in the same manner as previously stated.

\begin{table*}
\caption{Observed jet properties.}\label{jetproperties}
\begin{tabular}{lcccccccccc}
\hline
\hline

Name&nH ($10^{22}$&Photon&$\pm$&X-ray&$\pm$&X-ray Lum&$\pm$&8.46 GHz Radio&8.46 GHz Radio\\
&cm$^{-2}$)&Index&&Flux (Jy)&&(W Hz$^{-1}$)&&Flux (Jy)&Lum (W Hz$^{-1}$)\\

\hline

0313-192&$0.0318$&-&-&$4.29\times10^{-10}$&$2.47\times10^{-10}$&$4.55\times10^{15}$&$2.62\times10^{15}$&$0.0006$&$6.82\times10^{21}$\\
3C 15&$0.0303$&$2.63$&$0.20$&$1.34\times10^{-09}$&$5.64\times10^{-10}$&$1.70\times10^{16}$&$7.17\times10^{15}$&$0.1321$&$1.68\times10^{24}$\\
3C 296&$0.0186$&$1.96$&$0.24$&$1.52\times10^{-09}$&$5.37\times10^{-10}$&$1.89\times10^{15}$&$6.68\times10^{14}$&$0.0239$&$2.98\times10^{22}$\\
3C 31&$0.0541$&$2.22$&$0.09$&$9.55\times10^{-09}$&$9.34\times10^{-10}$&$5.82\times10^{15}$&$5.70\times10^{14}$&$0.0997$&$6.08\times10^{22}$\\
3C 346&$0.0561$&$2.06$&$0.26$&$9.58\times10^{-10}$&$3.88\times10^{-10}$&$6.67\times10^{16}$&$2.70\times10^{16}$&$0.1958$&$1.36\times10^{25}$\\
3C 371&$0.0484$&$2.01$&$0.04$&$3.90\times10^{-08}$&$1.89\times10^{-09}$&$2.34\times10^{17}$&$1.13\times10^{16}$&$0.0984$&$5.91\times10^{23}$\\
3C 465&$0.0506$&$1.85$&$0.44$&$1.65\times10^{-09}$&$5.09\times10^{-10}$&$3.16\times10^{15}$&$9.76\times10^{14}$&0.0841&$1.98\times10^{22}$\\
3C 66B&$0.0891$&$2.38$&$0.08$&$1.76\times10^{-08}$&$1.28\times10^{-09}$&$1.78\times10^{16}$&$1.30\times10^{15}$&$0.0841$&$8.52\times10^{22}$\\
3C 78&$0.0943$&$2.08$&$0.09$&$1.49\times10^{-08}$&$1.39\times10^{-09}$&$2.31\times10^{16}$&$2.16\times10^{15}$&$0.0981$&$1.52\times10^{23}$\\
3C 83.1&$0.1445$&-&-&$4.04\times10^{-10}$&$2.95\times10^{-10}$&$5.63\times10^{14}$&$4.12\times10^{14}$&$0.0181$&$2.53\times10^{22}$\\
3C 264&$0.0254$&$2.23$&$0.06$&$4.41\times10^{-08}$&$2.71\times10^{-09}$&$4.56\times10^{16}$&$2.80\times10^{15}$&$0.0889$&$9.20\times10^{22}$\\
4C 29.30&$0.0398$&-&-&$3.15\times10^{-09}$&$6.76\times10^{-10}$&$3.04\times10^{16}$&$6.52\times10^{15}$&$0.0304$&$2.94\times10^{23}$\\
B2 0206+35&$0.0604$&-&-&$3.18\times10^{-09}$&$8.45\times10^{-10}$&$9.75\times10^{15}$&$2.59\times10^{15}$&$0.0427$&$1.31\times10^{23}$\\
B2 0755+37&$0.0508$&$1.89$&$0.45$&$8.87\times10^{-09}$&$2.62\times10^{-09}$&$3.71\times10^{16}$&$1.10\times10^{16}$&$0.0316$&$1.32\times10^{23}$\\
Centaurus A&$0.0841$&$1.78$&$0.01$&$2.67\times10^{-07}$&$2.53\times10^{-09}$&$3.92\times10^{14}$&$3.71\times10^{12}$&$7.1413$&$1.05\times10^{22}$\\
M84&$0.0278$&$1.73$&$0.36$&$2.08\times10^{-09}$&$5.33\times10^{-10}$&$7.21\times10^{13}$&$1.84\times10^{13}$&$0.0328$&$1.13\times10^{21}$\\
M87&$0.0254$&$2.36$&$0.00$&$3.39\times10^{-07}$&$1.62\times10^{-09}$&$1.04\times10^{16}$&$4.96\times10^{13}$&$5.0457$&$1.55\times10^{23}$\\
NGC315&$0.0592$&$2.17$&$0.07$&$1.55\times10^{-08}$&$1.20\times10^{-09}$&$9.25\times10^{15}$&$7.14\times10^{14}$&$0.0841$&$5.02\times10^{22}$\\
NGC4261&$0.0158$&$2.21$&$0.13$&$4.27\times10^{-09}$&$4.78\times10^{-10}$&$4.91\times10^{14}$&$5.50\times10^{13}$&$0.0360$&$4.13\times10^{21}$\\
NGC6251&$0.0554$&$2.09$&$0.09$&$1.24\times10^{-08}$&$1.16\times10^{-09}$&$1.69\times10^{16}$&$1.59\times10^{15}$&$0.0347$&$4.75\times10^{22}$\\
PKS0851+202&$0.0302$&$1.75$&$0.05$&$2.74\times10^{-08}$&$1.69\times10^{-09}$&$8.14\times10^{18}$&$5.02\times10^{17}$&$0.0042$&$1.25\times10^{24}$\\
PKS2201+044&$0.0515$&$2.04$&$0.09$&$1.52\times10^{-08}$&$1.32\times10^{-09}$&$2.44\times10^{16}$&$2.13\times10^{15}$&$0.0408$&$6.57\times10^{22}$\\
S5 2007+777&$0.0879$&$1.25$&$0.17$&$4.35\times10^{-09}$&$8.84\times10^{-10}$&$1.68\times10^{18}$&$3.41\times10^{17}$&$0.0040$&$1.53\times10^{24}$\\

\hline

\end{tabular}

\vskip 5pt
\begin{minipage}{17.5cm}

`Name' lists the designation of the target as per the conventions described in Table 1. `nH' refers to the neutral hydrogen column density as given by the Colden calculator. Other columns are as described within the table, where $\pm$ indicates the uncertainty of the preceding column. Uncertainty in radio measurements are taken to be 3 per cent based on the uncertainty in absolute flux calibration of the VLA at the frequencies used.

\end{minipage}

\end{table*}

\begin{table*}
\caption{Jet volume and derived properties.}
\label{derivedproperties}
\begin{tabular}{lcccccccc}
\hline
\hline

Name&Volume&$\pm$&J(X)&$\pm$&J(8.46GHz)&$\pm$&X-ray to Radio&$\pm$\\
&(m$^3$)&&(W Hz$^{-1}$ m$^{-3}$)&&(W Hz$^{-1}$ m$^{-3}$)&&Flux Ratio\\

\hline

0313-192&$5.50\times 10^{59}$&$8.46\times 10^{59}$&$8.27\times 10^{-45}$&$1.36\times 10^{-44}$&$1.24\times 10^{-38}$&$1.90\times 10^{-38}$&$6.67\times 10^{-07}$&$3.85\times 10^{-07}$\\
3C 15&$2.89\times 10^{60}$&$1.91\times 10^{60}$&$5.90\times 10^{-45}$&$4.63\times 10^{-45}$&$5.82\times 10^{-37}$&$3.86\times 10^{-37}$&$1.01\times 10^{-08}$&$4.28\times 10^{-09}$\\
3C 296&$3.12\times 10^{59}$&$1.39\times 10^{59}$&$6.06\times 10^{-45}$&$3.45\times 10^{-45}$&$9.57\times 10^{-38}$&$4.28\times 10^{-38}$&$6.33\times 10^{-08}$&$2.25\times 10^{-08}$\\
3C 31&$3.43\times 10^{59}$&$1.80\times 10^{59}$&$1.70\times 10^{-44}$&$9.05\times 10^{-45}$&$1.77\times 10^{-37}$&$9.30\times 10^{-38}$&$9.58\times 10^{-08}$&$9.80\times 10^{-09}$\\
3C 346&$5.39\times 10^{60}$&$5.20\times 10^{60}$&$1.24\times 10^{-44}$&$1.30\times 10^{-44}$&$2.53\times 10^{-36}$&$2.44\times 10^{-36}$&$4.89\times 10^{-09}$&$1.99\times 10^{-09}$\\
3C 371&$1.48\times 10^{59}$&$1.06\times 10^{59}$&$1.58\times 10^{-42}$&$1.14\times 10^{-42}$&$3.99\times 10^{-36}$&$2.86\times 10^{-36}$&$3.97\times 10^{-07}$&$2.26\times 10^{-08}$\\
3C 465&$1.25\times 10^{60}$&$3.51\times 10^{59}$&$2.54\times 10^{-45}$&$1.06\times 10^{-45}$&$1.59\times 10^{-38}$&$4.50\times 10^{-39}$&$1.60\times 10^{-07}$&$4.96\times 10^{-08}$\\
3C 66B&$8.44\times 10^{58}$&$5.61\times 10^{58}$&$2.11\times 10^{-43}$&$1.41\times 10^{-43}$&$1.01\times 10^{-36}$&$6.71\times 10^{-37}$&$2.09\times 10^{-07}$&$1.65\times 10^{-08}$\\
3C 78&$1.62\times 10^{59}$&$6.70\times 10^{58}$&$1.42\times 10^{-43}$&$6.03\times 10^{-44}$&$9.39\times 10^{-37}$&$3.89\times 10^{-37}$&$1.52\times 10^{-07}$&$1.49\times 10^{-08}$\\
3C 83.1&$8.54\times 10^{59}$&$3.59\times 10^{59}$&$6.60\times 10^{-46}$&$5.57\times 10^{-46}$&$2.96\times 10^{-38}$&$1.25\times 10^{-38}$&$2.23\times 10^{-08}$&$1.63\times 10^{-08}$\\
3C 264&$3.21\times 10^{58}$&$2.13\times 10^{58}$&$1.42\times 10^{-42}$&$9.51\times 10^{-43}$&$2.87\times 10^{-36}$&$1.91\times 10^{-36}$&$4.96\times 10^{-07}$&$3.39\times 10^{-08}$\\
4C 29.30&$3.00\times 10^{61}$&$1.13\times 10^{61}$&$1.01\times 10^{-45}$&$4.40\times 10^{-46}$&$9.79\times 10^{-39}$&$3.71\times 10^{-39}$&$1.04\times 10^{-07}$&$2.24\times 10^{-08}$\\
B2 0206+35&$4.27\times 10^{59}$&$2.13\times 10^{59}$&$2.28\times 10^{-44}$&$1.29\times 10^{-44}$&$3.07\times 10^{-37}$&$1.53\times 10^{-37}$&$7.44\times 10^{-08}$&$1.99\times 10^{-08}$\\
B2 0755+37&$1.09\times 10^{60}$&$5.92\times 10^{59}$&$3.40\times 10^{-44}$&$2.10\times 10^{-44}$&$1.21\times 10^{-37}$&$6.60\times 10^{-38}$&$2.80\times 10^{-07}$&$8.33\times 10^{-08}$\\
Centaurus A&$1.09\times 10^{58}$&$1.12\times 10^{57}$&$3.60\times 10^{-44}$&$3.72\times 10^{-45}$&$9.61\times 10^{-37}$&$1.03\times 10^{-37}$&$3.74\times 10^{-08}$&$1.18\times 10^{-09}$\\
M84&$2.95\times 10^{56}$&$1.98\times 10^{56}$&$2.45\times 10^{-43}$&$1.76\times 10^{-43}$&$3.85\times 10^{-36}$&$2.59\times 10^{-36}$&$6.36\times 10^{-08}$&$1.64\times 10^{-08}$\\
M87&$2.11\times 10^{57}$&$1.12\times 10^{57}$&$4.93\times 10^{-42}$&$2.62\times 10^{-42}$&$7.33\times 10^{-35}$&$3.91\times 10^{-35}$&$6.72\times 10^{-08}$&$2.04\times 10^{-09}$\\
NGC315&$3.67\times 10^{59}$&$1.96\times 10^{59}$&$2.52\times 10^{-44}$&$1.36\times 10^{-44}$&$1.36\times 10^{-37}$&$7.28\times 10^{-38}$&$1.84\times 10^{-07}$&$1.53\times 10^{-08}$\\
NGC4261&$8.49\times 10^{58}$&$1.93\times 10^{58}$&$5.79\times 10^{-45}$&$1.47\times 10^{-45}$&$4.87\times 10^{-38}$&$1.12\times 10^{-38}$&$1.19\times 10^{-07}$&$1.38\times 10^{-08}$\\
NGC6251&$2.49\times 10^{60}$&$8.81\times 10^{59}$&$6.81\times 10^{-45}$&$2.50\times 10^{-45}$&$1.91\times 10^{-38}$&$6.80\times 10^{-39}$&$3.56\times 10^{-07}$&$3.50\times 10^{-08}$\\
PKS0851+202&$1.23\times 10^{64}$&$3.37\times 10^{63}$&$6.61\times 10^{-46}$&$1.85\times 10^{-46}$&$1.02\times 10^{-40}$&$2.80\times 10^{-41}$&$6.49\times 10^{-06}$&$4.46\times 10^{-07}$\\
PKS2201+044&$2.44\times 10^{60}$&$8.96\times 10^{59}$&$1.00\times 10^{-44}$&$3.77\times 10^{-45}$&$2.69\times 10^{-38}$&$9.91\times 10^{-39}$&$3.72\times 10^{-07}$&$3.43\times 10^{-08}$\\
S5 2007+777&$1.22\times 10^{64}$&$3.25\times 10^{63}$&$1.38\times 10^{-46}$&$4.63\times 10^{-47}$&$1.26\times 10^{-40}$&$3.38\times 10^{-41}$&$1.10\times 10^{-06}$&$2.25\times 10^{-07}$\\

\hline

\end{tabular}

\vskip 5pt
\begin{minipage}{17.5cm}

`Name' lists the designation of the target as per the conventions described in Table 1. `Volume' refers to that of the jet as modelled by a truncated cone.`J(...)' refers to the derived emissivity at the stated wavelength and `$\pm$' to the uncertainty of the preceding column.

\end{minipage}

\end{table*}

\begin{table*}
\caption{AGN core properties.}
\label{agnproperties}
\small
\begin{tabular}{lcccc}
\hline
\hline

Name&5 GHz Core&5 GHz Core&Core&References\\
&Flux(Jy)&Luminosity (W)&Prominence&\\

\hline

0313-192&0.1030&1.09E+24&-&-\\
3C 15&0.0297&3.78E+23&0.0177&-\\
3C 296&0.0770&9.59E+22&0.0376&1\\
3C 31&0.0920&5.61E+22&0.0442&1\\
3C 346&0.2200&1.53E+25&0.1583&1\\
3C 371&1.9807&1.19E+25&-&-\\
3C 465&0.2700&5.18E+23&0.0912&1\\
3C 66B&0.1820&1.84E+23&0.0619&1\\
3C 78&0.5780&8.99E+23&0.1533&-\\
3C 83.1&0.0400&5.58E+22&-&1\\
3C 264&0.2000&2.07E+23&0.1163&1\\
4C 29.30&0.1060&3.25E+23&-&2\\
B2 0206+35&0.0082&7.91E+22&-&2\\
B2 0755+37&0.1900&7.95E+23&0.1845&2\\
Cen A&5.5100&8.08E+21&0.0877&-\\
M84&0.1800&6.22E+21&0.0503&3\\
M87&4.0000&1.23E+23&0.0670&3\\
NGC315&0.4500&2.68E+23&0.3782&2\\
NGC4261&0.3036&3.49E+22&0.0742&-\\
NGC6251&0.8500&1.16E+24&-&3\\
PKS0851+202&1.7404&5.17E+26&0.6643&-\\
PKS2201+044&0.3189&5.13E+23&0.4304&-\\
S5 2007+777&0.9689&3.74E+26&-&-\\

\hline

\end{tabular}

\vskip 5pt
\begin{minipage}{9.5cm}

`Name' lists the designation of the target as per the conventions described in Table 1. `5 GHz Core Flux' values are taken from `References'. Where no reference is given, values have been measured directly from radio maps. Where a 5 GHz measurement does not exist, values have been extrapolated from other frequencies using a spectral index of 0. The references are as follows: 1, Hardcastle \& Worrall (1999); 2, Canosa \etal\ (1999); 3, Hardcastle \& Worrall (2000).

\end{minipage}

\end{table*}

Once the luminosity of a jet had been determined, it was then possible to derive a value for its volume emissivity. We again used DS9 to measure the width of the jet at the top and bottom of the selected region along with the total jet length. Radio maps were used for this purpose as they provided greater resolution than those obtained at X-ray wavelengths. Modelling the jet as a truncated cone, we then calculated its volume in the standard manner. The jet's luminosity was then divided by this value to give a value for its mean volume emissivity in the radio and X-ray wavebands.

As no previous studies have been carried out of this nature, as many properties as were practicable, given the available data, were tested for a correlation with those properties intrinsic to the X-ray jet. The data to which the X-ray properties were compared can be broadly broken down into two types; those which represent the host galaxy as a whole and those which are contained within the bound region of the X-ray jet. The properties of the host galaxies as a whole were the most straightforward to obtain as, thanks to works such as the two micron all sky survey (2MASS), apparent magnitudes at a given passband are widely available for a large number of nearby galaxies. Although the data were collated from previous studies, it was important for consistency and reliability that a limited number of trusted sources were used. As no single survey covers the entire frequency range in question, the Updated Zwicky catalogue (Emilio \etal\ 1999), the 2 mm all sky survey (Skrutskie \etal\ 2006) and the new catalogue of 4.85 GHz sources (Becker \etal\ 1991) were used to obtain the required values. A total of twelve properties were chosen for comparison against the X-ray jet luminosity and its photon index and values were retrieved from the NASA Extragalactic Database (NED)\footnote{http://nedwww.ipac.caltech.edu/}. For the AGN core luminosities, a well studied relation exists between emission at radio and X-ray wavelength (e.g. Hardcastle, Evans \& Croston 2006, 2009). We therefore used 5 GHz radio data taken from Hardcastle \& Worrall (1999), Canosa \etal\ (1999), Hardcastle \& Worrall (2000) and references therein as a measure of core luminosity. These values were also divided through by the total source 4.85 GHz luminosity to give its core prominence. This acts as a useful measure for comparison of properties and relations in which relativistic beaming may have a significant effect. A list of the chosen properties to be tested, and their origin, is shown in Table 2.

8.46 GHz radio measurements of the jet properties had to be obtained in a different way. Pre-existing radio maps were used which had been processed in the standard manner using AIPS and loaded into DS9. We then used the original region files created for determination of the X-ray flux to define the required radio jet area. We defined a representative background and used Funtools to calculate a background-subtracted measurement of the radio flux density. The luminosity of the radio jet was then derived from this value in the standard manner. For sources in which no 8.46 GHz data was available, measurements at alternative frequencies were made. These values were then converted using  $S_{8.46GHz} = S_\nu (8.46$ GHz$ / \nu)^{-\alpha}$ where $\nu$ is the measured frequency; we assume a value of $\alpha = 0.6$ to give an equivalent radio flux density at the required frequency of 8.46 GHz. Where no 5 GHz AGN core data were available in the references, these maps were also used to measure the core flux where a value of $\alpha = 0$ was assumed.

\subsection{Treatment of special cases}

The amount of information relating to specific target galaxies within the sample varies widely, both in terms of the raw data and the literature available. For the nearest bright sources, small-scale features can be resolved which in the fainter and more distant galaxies cannot be seen. We therefore had to consider carefully what information from detailed studies should be taken into account, bearing in mind the need for consistency across the sample.

One target which required such special consideration was that of Centaurus A. This is one of the best-studied galaxies within the sample and, at its distance of only $3.7$ Mpc, the jet can be very well resolved. Three issues had to be considered in this case. Firstly, we know that a dust lane cuts across the jet, creating greater than average extinction within this region. Accounting for the differences in column density along the length of the jet would help to improve the accuracy of the derived values; however, in the vast majority of the other target galaxies the much lower quality of the X-ray data means that it is not possible to produce a spatially resolved absorption map. As there is nothing to suggest that Centaurus A is in any way a special case in this sense, the increased absorption in this region was not accounted for but simply noted for later discussion. Secondly, it is known (e.g. Hardcastle \etal\ 2007) that the photon index along the jet varies in magnitude. However, as we are only considering the average value across the extent of the jet, and as any variation of the photon index cannot be determined to a comparable accuracy in the case of most galaxies within the sample, this was again ignored for current purposes. Finally, an uneven distribution of the background emission can be seen much more clearly in Centaurus A than in most other galaxies in the sample. The uneven distribution of background emission is not limited to the best-resolved galaxies, so our knowledge of this distribution could be used in determining the most appropriate background region, leading to a reduction in uncertainty of the X-ray flux and photon index without compromising the consistency of the data as a whole.

Until recently it was thought that extended radio sources, such as those found in FR-I galaxies, and their related X-ray components were a feature limited to elliptical galaxies. It has subsequently been shown by Keel \etal\ (2006) however, that this is not always true; the so-called `wrong galaxy' OL97 0313-192, has a spiral structure. Other examples of radio sources in disk type galaxies (e.g. Emont \etal\ 2009) have also been found, showing that this is not an isolated case; however, OL97 0313-192 is the only such example for which both \emph{Chandra} data and high resolution VLA maps are currently available. As this galaxy was already included on the XJET website and met all the selection criteria we saw no reason to reject this galaxy from consideration. This target also provided a good opportunity to test varying environments through which jets pass in relation to its elliptical counterparts.

\section{Results}
\label{results}

\begin{table*}
\caption{Correlated test results (full sample)}
\label{correlatedfull}
\begin{tabular}{llccc}
\hline
\hline
Tested Property&Jet Property&Spearman's Rank&$\tau / \sigma$&Significance\\
&&Coefficient&&Level\\
\hline

8.46 GHz Radio Emissivity&X-Ray Emissivity&0.87923&7.52691&1\%\\
8.46 GHz Radio Luminosity&X-Ray Luminosity  &0.87538&4.40877&1\%\\
4.85 GHz Radio Luminosity&X-Ray Luminosity  &0.75965&2.46778&1\%\\
Core 5 GHz&X-Ray Luminosity&0.74209&4.06009&1\%\\
Jet Volume&X-Ray Emissivity&0.72769&5.20254&1\%\\
Jet Volume&8.46 GHz Radio Emissivity&0.70923&4.20029&1\%\\
Core 5 GHz&8.46 GHz Radio Luminosity&0.58597&2.01274&1\%\\
Core 5 GHz&X-Ray Flux / 8.46 GHz Radio Flux&0.56917&2.85995&1\%\\
Core Prominence&X-Ray Flux / 8.46 GHz Radio Flux&0.71311&-&2\%\\
Absolute B Band Magnitude&X-Ray Luminosity&0.55038&2.28628&2\%\\

\hline
\end{tabular}
\vskip 5pt
\begin{minipage}{14cm}
Table of tested properties which show a statistically significant correlation. `Tested Property' lists the measured value compared to that shown in the adjacent `Jet Property' column. `8.46 GHz' refers to radio measurements confined to the same region as the X-ray jet as per Table 2. `Spearman's Rank' refers to the value of the Spearman's rank coefficient of the two properties. `$\tau / \sigma$' lists the value obtained from the partial correlation analysis when compared to redshift, where a value $>$2 indicates that the null hypothesis can be rejected at a 95 per cent confidence level. Photon index and core prominence are not expected to have any dependence on redshift hence, under the criteria detailed in Section 3.1, was excluded from a partial correlation analysis. `Significance Level' indicates to the significance level at which the relation is found.
\end{minipage}
\end{table*}

\subsection{Statistical analysis}

We searched for correlations between the measured values using a Spearman's rank test. If, and only if, a correlation was present between two quantities that both depended on redshift (e.g. two luminosities or magnitudes) then we used a partial correlation analysis based on the methodology presented by Akritas \& Siebert (1996). These partial correlation results were then used to determine whether the relation was still statistically significant to a 5 per cent level. Statistically significant correlations for the full sample are listed in Table 6 and for the same restricted to \emph{bona fide} FR-Is in Table 7. Correlations not found to be significant for the full sample are listed in in Table 8 and for the restricted sample in Table 9.

\begin{table*}
\caption{Correlated test results (\emph{bona fide} FR-Is only)}
\label{correleatedFRIs}
\begin{tabular}{llccc}
\hline
\hline

Tested Property&Jet Property&Spearman's Rank&$\tau / \sigma$&Significance\\
&&Coefficient&&Level\\

\hline

8.46 GHz Radio Emissivity&X-ray Emissivity&0.89461&6.27738&1\%\\
8.46 GHz Radio Luminosity  &X-Ray Luminosity  &0.87010&8.25512&1\%\\
Jet Volume&8.46 GHz Radio Emissivity&0.75490&3.62678&1\%\\
Core 5 GHz&X-Ray Flux / 8.46 GHz Radio Flux &0.74706&2.70024&1\%\\
Core 5 GHz&X-Ray Luminosity  &0.67353&2.68152&1\%\\
Core Prominence&X-Ray Flux / 8.46 GHz Radio Flux &0.69780&-&2\%\\
8.46 GHz Radio Luminosity  &Photon Index &0.64835&-&2\%\\
Jet Volume&X-Ray Emissivity&0.64216&3.67540&2\%\\
X-Ray Luminosity  &Photon Index &0.60440&-&5\%\\
Core 5 GHz&8.46 GHz Radio Luminosity  &0.60000&2.50615&5\%\\

\hline

\end{tabular}

\vskip 5pt

\begin{minipage}{14cm}

`Tested Property' lists the measured value compared to that shown in the adjacent `Jet Property' column. `8.46 GHz' refers to radio measurements confined to the same region as the X-ray jet as per Table 2. `Spearman's Rank' refers to the value of the Spearman's rank coefficient of the two properties. `Significance Level' indicates the significance level at which the relation is found.

\end{minipage}

\end{table*}

\subsection{Photon index relations}
We see from Table 6 that when the full sample is considered, no properties are found to be related to the photon index of the jets; However, in our subsample of \emph{bona fide} FR-Is we find relations with both the X-ray and 8.46 GHz radio luminosity.

We see that in both of these relations photon index increases with increasing luminosity (Figure 2); relative intensity of high-energy emission is reduced in more luminous sources. It is evident from the plots that both these relations can be described in the form of a power law. Using the least squares fit method in the standard manner, we find that the relation to the X-ray luminosity is given by $\Gamma \propto L_{X}^{0.03 \pm 0.01}$ where $\Gamma$ is the photon index. In the same manner we find that the relation to the 8.46 GHz radio luminosity is given by $\Gamma \propto L_{R8.46}^{0.04 \pm 0.02}$.

Given the result of Hardcastle \etal\ (2007) on the variation of photon index with distance along the jet in Centaurus A, we searched for a correlation between photon index and jet length in our sample; however, it is immediately evident from the values in Table 9 that no correlation is present. 

We also searched for correlations between the photon index and 9 other intrinsic properties of the host galaxy; however, no significant correlations were found in either the full sample, or the subsample (Tables 8 \& 9).

\subsection{Derived ratios}

As discussed in the introduction, particles emitting at X-ray wavelengths must be accelerated \emph{in situ}. The efficiency with which particles residing at low energies are accelerated is therefore well represented by the ratio $L_{X}/L_{R8.46}$, which is confined to the region of the X-ray jet. We see from the results of the Spearman's rank test that the ratio of $L_{X} / L_{R8.46}$  is related to the 5 GHz radio core luminosity and core prominence. We discuss this relation further in the next section. Outside of these relations we observe no link between the derived ratios and the properties of the host galaxy.

This lack of correlation between $L_{X}/L_{R8.46}$ and luminosities at X-ray and radio wavelengths does, however, provide useful information about the processes involved. From the plot of $L_{X} / L_{R8.46}$ against 8.46 GHz radio luminosity, as shown in Figure 2, we find that the acceleration efficiency as measured by $L_{X} / L_{R8.46}$ shows little dependence on luminosity over a wide range of $L_{R8.36}$ if \emph{bona fide} FR-Is are considered. This is again discussed further in Section 4.2.

\subsection{Luminosity \& emissivity relations}

We would expect radio and X-ray luminosities to be correlated, and we see from the results in Tables 6 and 7 that this is indeed the case for both the full and restricted sample; there is a strong correlation in the Spearman's rank test. This relationship is at the 1 per cent significance level in both cases and so we can be fairly confident that a true correlation exists; however, the emissivity's lack of correlation to the photon index is more unexpected and opens up an additional mystery which we discuss in Section 4. It is not surprising that this relation is found in both of the samples as the targets excluded from the restricted sample are primarily high luminosity galaxies at both X-rays and radio wavelengths (red points, Figure 2). This therefore precludes any definite inferences from being drawn from the full sample as even if a break or multiple relations are present, these additional galaxies are small enough in number and high enough in power at both wavelengths that they will still score well in a Spearman's rank test. Looking at Figure 2, we can see evidence of this with the much greater scatter exhibited by these high power objects compared to the \emph{bona fide} FR-Is.

For the restricted sample, where the correlation is better constrained, we see that the relations between X-ray luminosity and emissivity to those at 8.46 GHz wavelengths can again be described by a power law. This strong correlation, discussed further in Sections 3.4 and 4.2, lends the first good empirical support to the previously assumed relation between these quantities. It also provides reinforcement to the relationships found for \emph{bona fide} FR-Is with the photon index, as a correlation between the two luminosities naturally leads to an expectation that either both, or neither, should be correlated with any third variable.

We also see that volume emissivity reduces as a function of volume at both X-ray and radio wavelengths in both samples. The lack of correlation between volume emissivity and luminosity, however, implies that the effects of an increasing volume outweighs the effects of a reduced emissivity in terms of the total power output of the jet.

Relationships in both samples are also seen to exist between X-ray luminosity of the jet and the 5 GHz radio core luminosity (Figure 2). We know that both of these properties depend on both jet power and orientation, but only the X-ray jet has a dependence on acceleration efficiency between X-ray and radio energies. The ratio of $L_{X}/L_{R8.46}$ which, as discussed in Section 3.3, can be taken as a measure of acceleration efficiency, also has a dependence on orientation. This is because as the jets are viewed from smaller angles to the line of sight we expect to observe an increase in the Doppler factor. Assuming a broken power law or similar concave spectrum, between emission at radio and X-ray wavelengths, the increase in emission at X-ray wavelengths will be greater due to the steeper spectral index at these energies compared to the relatively flat spectral index of the radio emission. This leads to an apparent increase in the ratio of $L_{X}/L_{R8.46}$. Knowing these dependencies, we can compare the 5 GHz core luminosity to $L_{X}/L_{R8.46}$. As out of these two properties only the core luminosity is dependent on power, so if we observe a relation is either not present or much worse than that between X-ray luminosity and the 5 GHz radio core, it is likely that the jet power is the dominant property. However, if we find that the relation is comparable or has improved, it is the common dependence on orientation and hence beaming that is the most likely cause of the relation. From Tables 6 and 7 we clearly see that in both samples relationships are found to comparable significance.

As mentioned in Section 2.2, core prominence can be used as a measure of beaming within these galaxies. We therefore decided to test this property against the ratio of $L_{X}/L_{R8.46}$ as additional confirmation of this conclusion. As expected we see that a relation at the 2 per cent level is present. This drop in significance is not particularly surprising, as although core prominence is a commonly used measure of beaming, it does not allow for factors such as ageing and so additional scatter is introduced. We therefore believe that beaming rather than a relationship with the intrinsic power of the jet is the most likely explanation for this relationship in both cases.

There are two X-ray luminosity relations that exist in the full sample, which are absent from the restricted sample, being those with the absolute B band magnitude and 4.85 GHz radio luminosity. The relation to B band magnitude we believe is again a result of the effects of orientation. The B-band luminosities of BL Lac objects are, by definition, partly or wholly related to their nuclear jets. At the same time beaming causes an increase in X-ray luminosity, so for the full sample which contains several BL Lacs, the inclusion of these objects will lead to a natural correlation between B band and X-ray luminosities.

We believe that the correlation between the 4.85 GHz radio luminosity and the X-ray luminosity is also due to the nature of the full sample over that of the \emph{bona fide} FR-Is. The jets and lobes of the intermediate FR-I / II and BL Lacs included in the full sample are observed to have a much higher power at radio wavelengths than their FR-I counter parts (Figure 2). The radio emission of these jets is included in the measurement of the galaxy's total 4.85 GHz luminosity and so, for these galaxies, the jets are likely to dominate the total galactic emission. We therefore believe that this relation is again due to sample selection, rather than to any intrinsic property of the galaxies.

Although some correlations were initially seen to exist in other host galaxy properties when subjected to the Spearman's rank test, we found upon applying partial correlation analysis that these were merely a result of redshift (Tables 8 and 9). We therefore conclude that there are no statistically significant relations linking the properties of the host galaxy to that of the jet.

\begin{table*}
\caption{Uncorrelated test results (full sample)}
\label{resultstablenoncorr}
\begin{tabular}{llcc}
\hline
\hline
Tested Property&Jet Property&Spearman's Rank&$\tau / \sigma$\\
&&Coefficient&\\
\hline

Absolute K Band Magnitude &X-Ray Luminosity  &0.48271&1.63588\\
X-Ray Luminosity  &X-Ray Flux / 8.46 GHz Radio Flux &0.45923&1.77570\\
Absolute J Band Magnitude &X-Ray Luminosity  &0.40902&1.21711\\
8.46 GHz Radio Emissivity&Photon Index&0.43609&-\\
X-ray Emissivity&Photon Index&0.37744&-\\
Absolute H Band Magnitude &X-Ray Luminosity  &0.31880&-\\
Core 5 GHz&Photon Index&0.25614&-\\
8.46 GHz Radio Luminosity  &Photon Index &0.22556&-\\
X-Ray Flux / 8.46 GHz Radio Flux &Photon Index&0.20902&-\\
Absolute H Band Magnitude &X-Ray Flux / 8.46 GHz Radio Flux &0.19549&-\\
Core 5 GHz&X-ray Emissivity&0.17095&-\\
Absolute B Band Magnitude &Photon Index &0.15196&-\\
4.85 GHz Radio Luminosity   &Photon Index &0.14551&-\\
Absolute J Band Magnitude &X-Ray Flux / 8.46 GHz Radio Flux &0.09323&-\\
4.85 GHz Radio Luminosity   &X-Ray Flux / 8.46 GHz Radio Flux &0.08596&-\\
Absolute B Band Magnitude &X-Ray Flux / 8.46 GHz Radio Flux &0.06316&-\\
8.46 GHz Radio Luminosity  &X-Ray Flux / 8.46 GHz Radio Flux &0.05692&-\\
Absolute K Band Magnitude &Photon Index &0.04231&-\\
Absolute J Band Magnitude &Photon Index &0.04025&-\\
X-Ray Luminosity  &Photon Index &0.03158&-\\
Absolute K Band Magnitude &X-Ray Flux / 8.46 GHz Radio Flux &0.01955&-\\
Absolute H Band Magnitude &Photon Index &0.00310&-\\

\hline
\end{tabular}
\vskip 5pt
\begin{minipage}{14cm}
Table of tested properties which do not show a statistically significant correlation. `Tested Property' lists the variable compared to that shown adjacent as `Jet Property'. `8.46 GHz' refers to radio measurements confined to the region of the X-ray jet and `4.85 GHz' to total emission from the jet and host galaxy as per Table 2. `$\tau / \sigma$' lists the value obtained from the partial correlation analysis when compared to redshift, where a value $>$2 indicates that the null hypothesis can be rejected at a 95 per cent confidence level. Note that luminosity is not tested against jet volume, as this is naturally biased towards a correlation due to the selection criteria. `Spearman's Rank' refers to the value of the Spearman's rank coefficient of the two properties.
\end{minipage}
\end{table*}

\begin{table*}
\caption{Uncorrelated test results (\emph{bona fide} FR-Is only)}
\label{noncorrFRIs}
\begin{tabular}{llcc}
\hline
\hline
Tested Property&Jet Property&Spearman's Rank&$\tau / \sigma$\\
&&Coefficient&\\
\hline

4.85 GHz Radio Luminosity   &X-Ray Luminosity  &0.68352&1.95037\\
X-Ray Luminosity  &X-Ray Flux / 8.46 GHz Radio Flux &0.57108&1.79649\\
Absolute K Band Magnitude &X-Ray Luminosity  &0.55604&1.46254\\
Absolute J Band Magnitude &X-Ray Luminosity  &0.55604&1.46254\\
Absolute H Band Magnitude &X-Ray Luminosity  &0.55604&1.02397\\
Absolute B Band Magnitude &X-Ray Luminosity  &0.42143&-\\
X-ray Emissivity&Photon Index&0.37143&-\\
8.46 GHz Radio Emissivity&Photon Index&0.35824&-\\
X-Ray Flux / 8.46 GHz Radio Flux &Photon Index &0.31868&-\\
4.85 GHz Radio Luminosity   &X-Ray Flux / 8.46 GHz Radio Flux &0.30549&-\\
4.85 GHz Radio Luminosity   &Photon Index &0.29670&-\\
Absolute H Band Magnitude &Photon Index &0.18681&-\\
Absolute J Band Magnitude &Photon Index &0.18681&-\\
Absolute K Band Magnitude &Photon Index &0.18681&-\\
8.46 GHz Radio Luminosity  &X-Ray Flux / 8.46 GHz Radio Flux &0.14951&-\\
Absolute H Band Magnitude &X-Ray Flux / 8.46 GHz Radio Flux &0.11071&-\\
Absolute J Band Magnitude &X-Ray Flux / 8.46 GHz Radio Flux &0.11071&-\\
Absolute K Band Magnitude &X-Ray Flux / 8.46 GHz Radio Flux &0.11071&-\\
Core 5 GHz&Photon Index&0.10989&-\\
Core 5 GHz&X-ray Emissivity&0.07941&-\\
Absolute B Band Magnitude &X-Ray Flux / 8.46 GHz Radio Flux &0.02647&-\\
Absolute B Band Magnitude &Photon Index &0.01648&-\\

\hline
\end{tabular}
\vskip 5pt
\begin{minipage}{14cm}
Table of tested properties which do not show a statistically significant correlation. `Tested Property' lists the variable compared to that shown adjacent as `Jet Property'. `8.46 GHz' refers to radio measurements confined to the region of the X-ray jet and `4.85 GHz' to total emission from the jet and host galaxy as per Table 2. `$\tau / \sigma$' lists the value obtained from the partial correlation analysis when compared to redshift, where a value $>$2 indicates that the null hypothesis can be rejected at a 95 per cent confidence level. Note that luminosity is not tested against jet volume, as this is naturally biased towards a correlation due to the selection criteria. `Spearman's Rank' refers to the value of the Spearman's rank coefficient of the two properties.
\end{minipage}
\end{table*}

\section{Discussion}
\label{discussion}

\subsection{Current models of particle acceleration}

It is generally agreed (e.g. Hardcastle \etal\ 2003; Kataoka \etal\ 2006) that at least two forms of particle acceleration are present within the jets of radio galaxies. There is growing evidence, particularly in the case of Centaurus A (e.g. Goodger \etal\ 2009), that emission relating to small scale features such as knots is due to shock acceleration caused by interaction with a compact stationary or slow moving body. This acceleration is seen to dominate emissions from the innermost sections of the jet in galaxies such as Centaurus A (e.g. Hardcastle 2007); however, the particles responsible for the more diffuse emission which dominates the outer jet regions cannot be explained in the same manner, so the particles radiating at these wavelengths must be accelerated by a second process (Hardcastle \etal\ 2004a; Kataoka \etal\ 2006). Our inability to resolve all but the closest jets to this level of detail means it is uncertain whether this is the case for all FR-I galaxies; however, it is likely that diffuse emission will be present in the majority, if not all cases, based on those objects in which they can currently be resolved.

\subsection{Key emission process variables}

Determining which, if any, of the processes discussed in Section 4.1 causes the acceleration of particles to the very high energies required for X-ray emission to be produced has proved to be difficult, and a definitive answer has unsurprisingly not emerged in this paper; however, the results of Section 3.2 do carry significant implications for key features of the acceleration process.

The photon index gives a measure of how emission intensity changes as a function of frequency; therefore it can be interpreted as giving an indication of how efficiently particles are being accelerated at the very highest energies, as it effectively tells us about the ratio between the number of low-energy and high-energy particles producing the X-ray emission. A lower photon index represents a harder X-ray distribution, with a greater proportion of particles having been accelerated to higher energies. We see that for \emph{bona fide} FR-Is there is a reduction of the relative intensity of high-energy emission in more luminous sources. This suggests that acceleration to the highest energies is either less efficient, or that particles suffer greater radiative losses in conditions in which the jet luminosity is highest.

However, the photon index is not the only measure of acceleration efficiency. Whereas the photon index represents the efficiency of particle acceleration to energies required for the hardest X-ray emission, the ratio of $L_{X}/L_{R8.46}$ provides a means of determining efficiency in terms of particles being accelerated from the energies required for radio emission, to those required for X-ray emission. It is interesting to note therefore that the BL Lac and intermediate FR-I / II objects occupy a range in this quantity almost double that in which the \emph{bona fide} FR-Is reside. This increased scatter is a recurring feature of all common relationships between the full and restricted samples, suggesting that we are dealing with at a minimum two separate populations. For BL Lac objects, which are thought to be beamed FR-I galaxies, this is likely to be a result of beaming effects which increase greatly at small jet angles to the line of sight. However, the greatest scatter is exhibited in the intermediate objects which lie between the FR-I and FR-II classifications. We therefore suggest that this is a second population, operating under a separate set of conditions or mechanisms to that of the \emph{bona fide} FR-I subsample.

Considering only the \emph{bona fide} FR-Is in Figure 2, the ratio of $L_{X}$ and $L_{R8.46}$ of the subsample falls within two orders of magnitude, and is reasonably constant across the range of photon indices. If the full sample does indeed contain two separate populations and so can be excluded from consideration, the relationship between X-ray and radio luminosities becomes near linear as $(L_{X}) \propto  L_{R8.46}^{1.11 \pm 0.21}$.  As the value of $L_{X}/L_{R8.46}$ also represents the relation between X-ray and radio luminosities with the power law removed, the plot of $L_{X}/L_{R8.46}$ against $L_{R8.46}$ therefore simply represents the scatter around this relationship. A linear scaling rather than a power law therefore provides the best description of the luminosity-luminosity correlation for bona fide FR-Is, providing a consistent interpretation across all observed data.

The lack of correlation between $L_{X}/L_{R8.46}$ and photon index also holds implications for particle acceleration models. As the two efficiencies are seen to be independent of each other, any models explaining the dependence of photon index on jet luminosity must also ensure they do not also affect the overall observed acceleration efficiency. In Section 4.3 we test one such model in which radiative losses dominate the relative number of particles emitting at the highest energies.

We also see from Section 3.2 and 3.3 that the host galaxy luminosity in various optical and infrared bands, and hence the average host galaxy environment in which the jets reside, have little to no impact on either the luminosity or photon index of the X-ray jet in either of the samples. If the host galaxy environment did have a strong effect on the jet luminosities, we would most likely notice a peculiarity in the properties of the spiral galaxy OL97 0313-192 due to its morphology and composition in comparison to the ellipticals which comprise the majority of the sample. However, we find in fact that this is far from the case; in fact the source exhibits very average values both in terms of X-ray and radio jet luminosities, and is indistinguishable on our plots from the elliptical galaxies. The same reasoning can also be applied to the photon index, in that the ability to accelerate particles to the high-energy end of the X-ray range seems to be independent of the properties of the host galaxy through which the jet passes. Again, if the region of the host galaxy through which the jet passes were a dominant factor in determining the acceleration to the highest electron energies, the photon index in the environment of the spiral galaxy would be unlikely to take such an typical value. Therefore the mechanism causing the acceleration of these particles must on these large scales, be largely independent of the host galaxy's bulk properties. Although the case of OL97 0313-192 strengthens the case for this low dependence, we emphasise that the key observation underpinning this weak dependence on the host galaxy's bulk properties is the lack of correlation observed between jet properties and the host galaxy's luminosity in various optical and infrared bands.

\begin{figure*}
\centering
\includegraphics[angle=270,width=8cm]{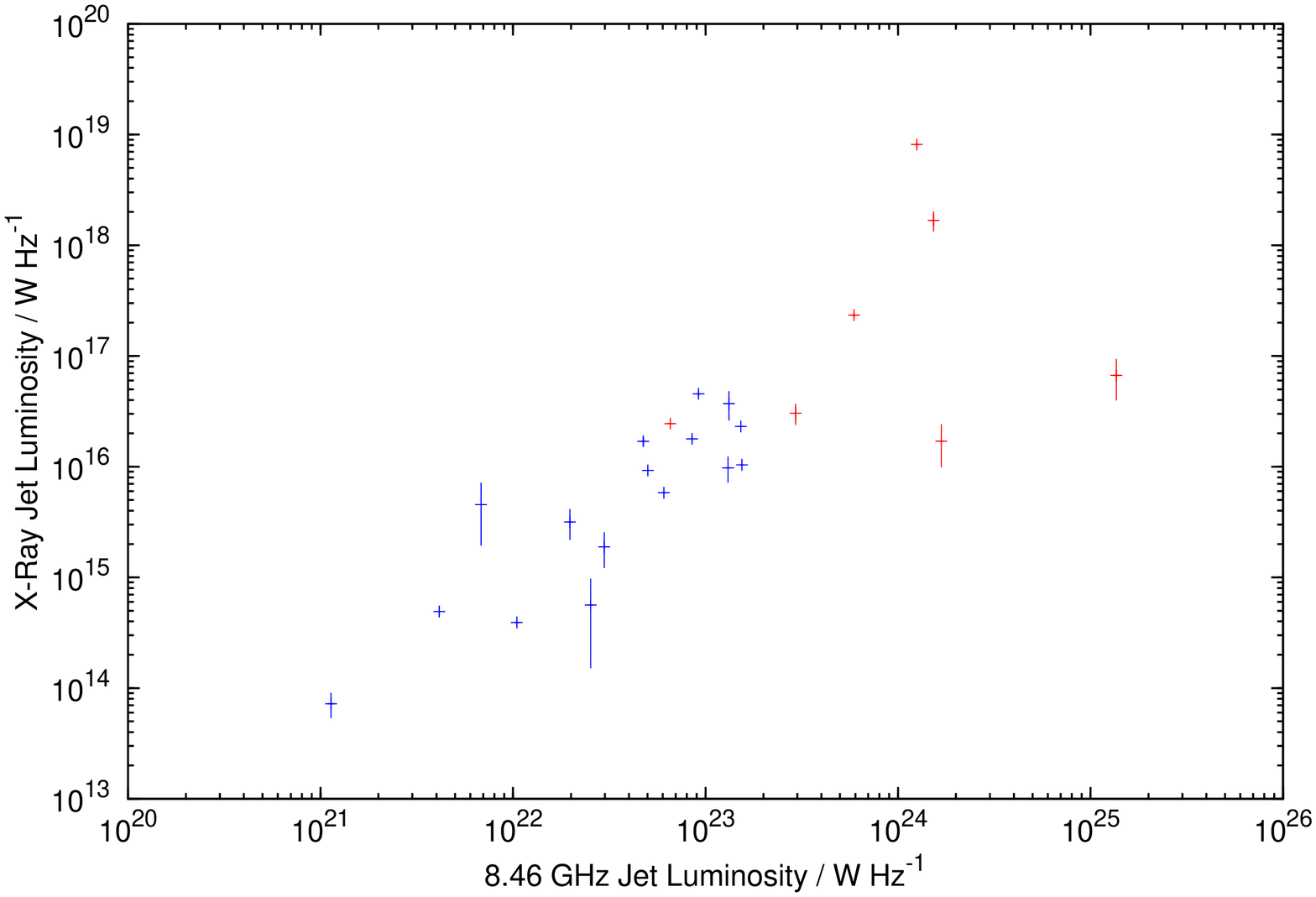}
\includegraphics[angle=270,width=8cm]{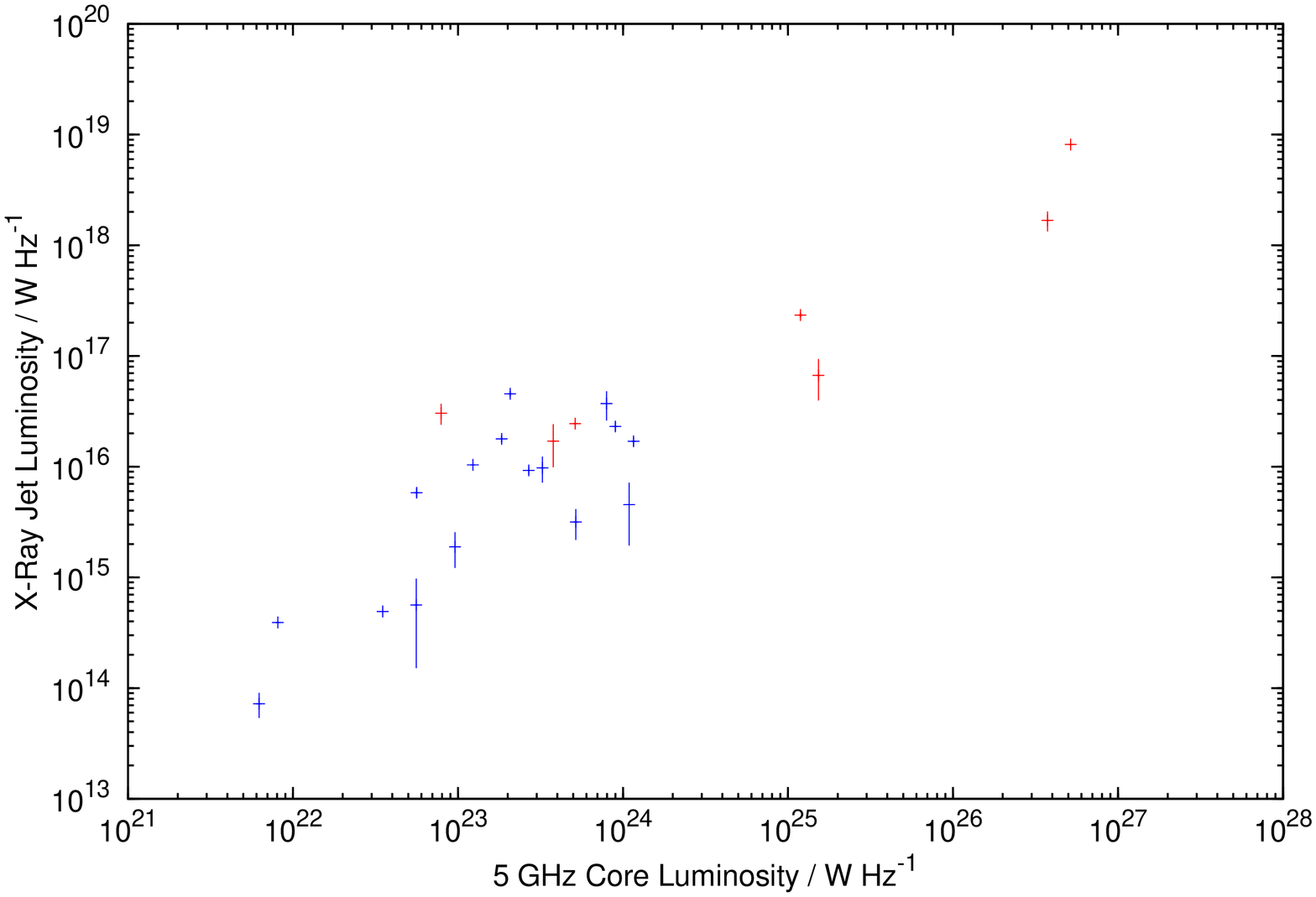}\\
\includegraphics[angle=270,width=8cm]{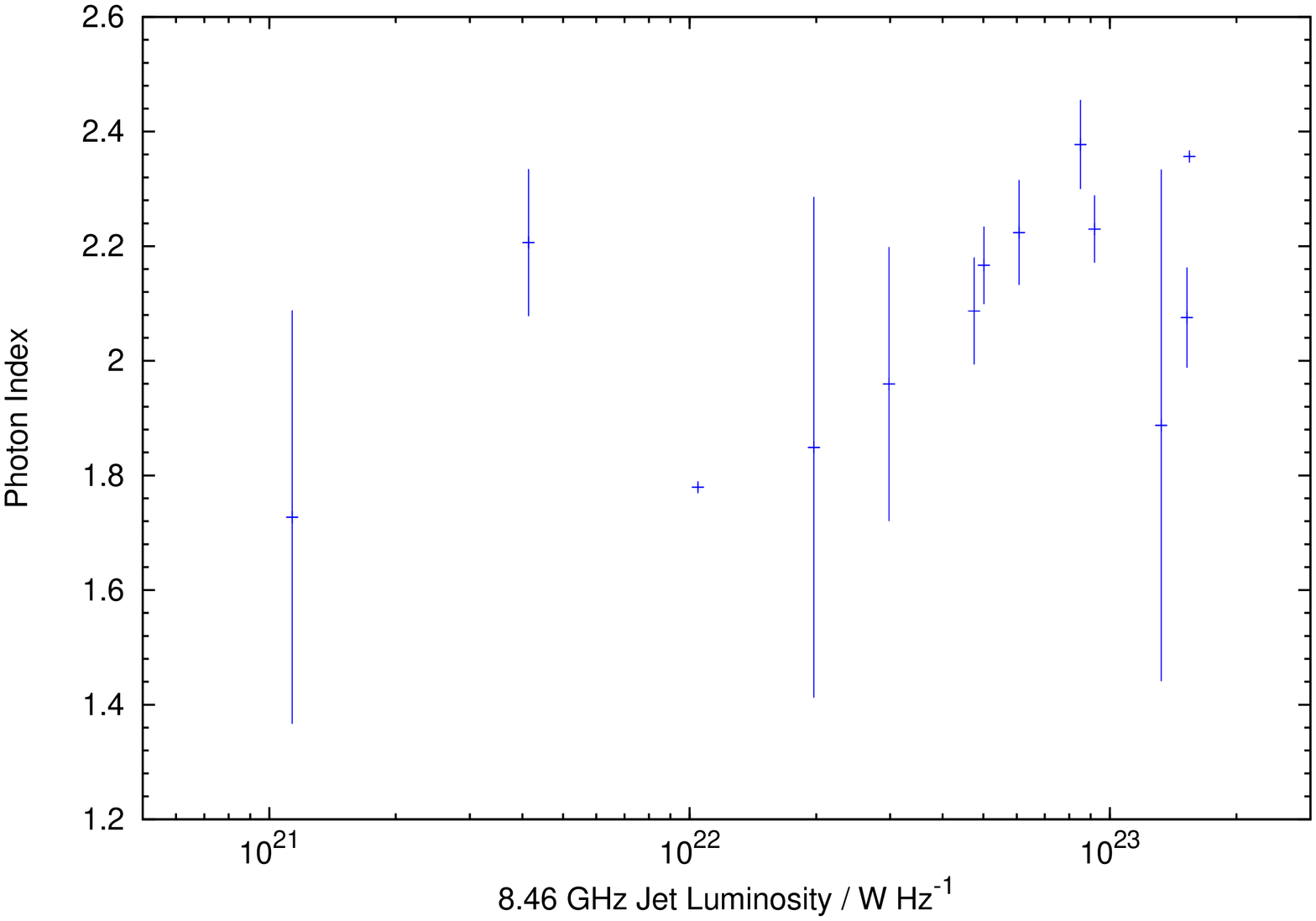}
\includegraphics[angle=270,width=8cm]{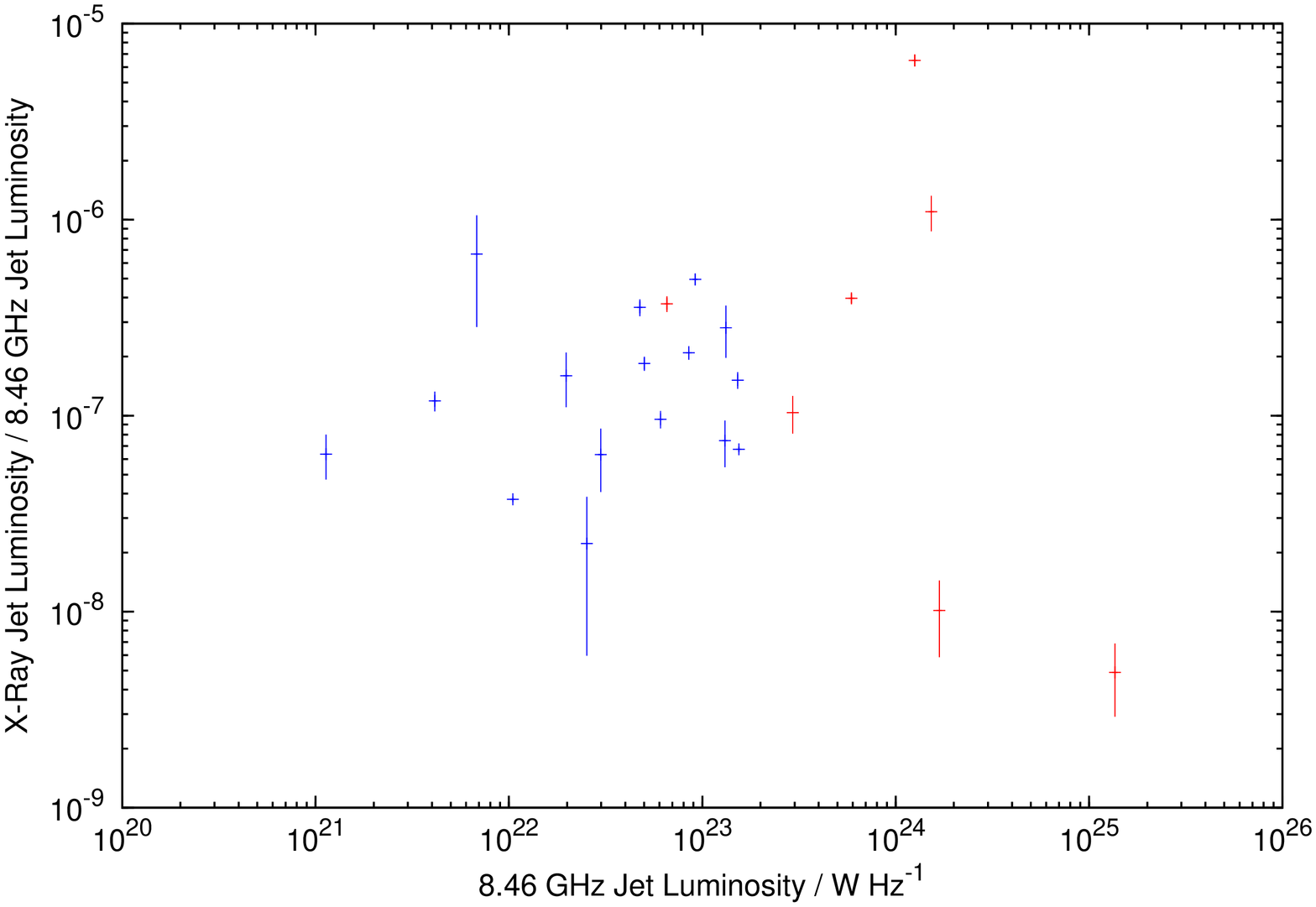}\\
\caption{Top left: The strong correlation between jet luminosity at radio and X-ray wavelengths. Top right: The correlations between 5 GHz core luminosity and X-ray jet luminosity discussed in Section 3.4. Bottom left: The relation between 8.46 GHz radio luminosity and photon index in \emph{bona fide} FR-Is discussed in Section 3.2. Bottom right: Plot of flux ratio against radio luminosity discussed in Section 4.2.
\newline Blue points are those classed as \emph{bona fide} FR-Is. Red points represent the BL Lacs and intermediate cases included in the full sample.}
\label{statsigresultplots}
\end{figure*}

\subsection{Testing of viable acceleration models}

As mentioned in Section 3.2, the photon index relations must either be a result of decreased acceleration efficiency, or increased radiative losses, at higher jet luminosities. Although it has been established that for at least some compact jet regions inverse-Compton losses dominate (e.g. Perlman \etal\ 2011), synchrotron losses will dominate on larger scales, and so the properties of a jet should depend on its magnetic field strength, in the sense that jets with a higher field strength will suffer stronger synchrotron losses and should therefore be less able to accelerate particles to the highest energies. In this section we investigate whether these losses alone can give rise to the photon index relation we observe in the sample of \emph{bona fide} FR-Is in a similar manner to the process proposed for the emission from the hotspots of FR-II sources (e.g. Hardcastle \etal\ 2004b)

For synchrotron emission, intensity is peaked around a critical frequency which scales as \begin{equation}\label{critfreq}\nu_{crit} \propto \gamma^{2}B\end{equation} However, the effect of particle energy loss also counteracts this effect, scaling as \begin{equation}\label{losses} \frac {{\rm d}E}{{\rm d}t} \propto \gamma^{2}B^{2}\end{equation} The loss timescale of an emitting particle is given by \begin{equation}\label{tau}\tau = \frac {E}{{\rm d}E/{\rm d}t} \end{equation} where $E=\gamma mc^{2}$ is the particle energy. Applying the known scaling for synchrotron losses (\ref{losses}) to equation (\ref{tau}) we therefore find that \begin{equation}\label{taufull}\tau \propto \frac {\gamma mc^{2}}{\gamma^{2} B^{2}}\end{equation} Substituting for $\gamma$ in terms of the critical frequency in equation (\ref{critfreq}) we find that \begin{equation}\label{taubfield}\tau \propto \frac {mc^{2}}{\nu^{1/2} B^{3/2}}\end{equation} As for our observations we can take the observing frequency $\nu$ to be fixed, we therefore find that the energy time loss scales as \begin{equation}\label{tauscales}\tau \propto \frac {1}{B^{3/2}}\end{equation}

Although values for the magnetic field within the jets cannot currently be measured directly, the field strength within the jets is often assumed to be around the equipartition value (e.g. Leahy 1991) in which the energy contribution of the field is equal to that of the emitting particles. Under this assumption, luminosity might be expected to increase with increasing field strength; therefore, if the magnetic field strength were the dominant factor in determining the jet properties, the short decay timescales required to describe the $\Gamma \propto L$ relations observed in our results could be explained due to a greater dependence of losses on the $B$-field. A model in which the $B$-field increases with luminosity, thus giving rise to shorter loss timescales, could explain the observed photon index softening seen in our results for an increasing luminosity. 

However, there is a problem with this model. For equipartition, the local magnetic field strength in fact also depends on emissivity, as $J(\nu) \propto N_0 \nu^{-(p-1)/2} B^{(p+1)/2}$ (e.g. Longair 1994; Bordovitsyn 1999). We would therefore expect to see emissivity scale with photon index if the $B$-field is in equipartition. The results presented in Table 6 show that this is not the case. There is also strong evidence from the observations of Centaurus A (Hardcastle \etal\ 2007) that losses in this manner do not dominate the emission process. They show that there is significant variation in the values of the photon index along the length of the jet which are hard to explain in the context of a $B$-field dominated model since presumably, if anything, the $B$-field is decreasing along the jet whilst $\Gamma$ is increasing.

It is therefore unlikely that the model as it stands provides a good description of the true nature of the jets and we are therefore left with two viable options. The first is that the $B$ dominated model is incorrect and losses due to the large scale magnetic fields do not play a significant role in the process that sets the observed $\Gamma$; the second is that the magnetic field does not lie in equipartition throughout the jet, meaning that the mean volume emissivity is not representative of the field strength.

One thing that is clear is that the models of particle acceleration currently proposed do not account for the bulk relations that we have found. If a true description of the acceleration process is to be made, any proposed model must account for these general relations, as well as those within small scale features and individual galaxies.

\section{Conclusions}

Although no definitive conclusions can yet be drawn about the cause of particle acceleration, we have found a number of key relations which are not currently considered in models of FR-I galaxies, along with empirical evidence for relationships which have been long assumed to exist. These can be summarized as follows;

\begin{enumerate}

\item We find that the luminosity of X-ray jets scales linearly with luminosity at radio wavelengths.

\item We find that the interstellar medium through which a jet passes is unlikely to have a direct effect on determining the intrinsic acceleration properties of jets.

\item We find that for \emph{bona fide} FR-Is, the photon index is related to jet luminosity at both X-ray and radio wavelengths, but shows no correlation to any other tested property.

\end{enumerate}

Although we have presented the first large sample investigation of the properties of X-ray jets in FR-Is, the uncertainties involved are limited by the number of FR-Is known to have X-ray jets and the amount of data available for those which do. We therefore suggest that future attempts should be made to verify and constrain the relations found within this paper. Increasing the sample size as more FR-I galaxies with suitably resolvable X-ray jets are discovered will help reduce the uncertainties involved. Even within the constraints imposed by the available data, it is apparent that many previously unknown correlations exist.

We have tested a simple model in which large-scale magnetic field variations are primarily responsible for determining jet properties; however, we found that this model is inconsistent with our best estimates of the relative magnetic field strength in our sample. If an accurate description of FR-I jets is to be achieved, it is vital that these relations should be accounted for within future models.

\section*{Acknowledgements}

We wish to thank William Keel for providing the radio map of OL97 0313-192, making a full analysis of this unusual galaxy possible.

JJH thanks the University of Hertfordshire for a studentship and MJH the Royal Society for generous financial support through the University Research Fellowships scheme.

We also wish to thank the referee, Eric Perlman, for his constructive comments and suggestions.

\end{document}